# PINK: physical-informed machine learning for lattice thermal conductivity

**Yujie Liu, Xiaoying Wang, Yuzhou Hao, Xuejie Li, Jun Sun, Turab Lookman, Xiangdong Ding, Zhibin Gao***

State Key Laboratory for Mechanical Behavior of Materials, School of Materials Science and Engineering, Xi'an Jiaotong University, Xi'an 710049, Shaanxi, China.

***Correspondence to:** Prof. Zhibin Gao, State Key Laboratory for Mechanical Behavior of Materials, School of Materials Science and Engineering, Xi'an Jiaotong University, No.28, West Xianning Road, Xi'an 710049, Shaanxi, China. E-mail: zhibin.gao@xjtu.edu.cn

## Abstract

Lattice thermal conductivity ($\kappa_L$) is crucial for e icient thermal management in electronics and energy conversion technologies. Traditional methods for predicting $\kappa_L$ are often computationally expensive, limiting their scalability for large-scale material screening. Empirical models, such as the Slack model, offer faster alternatives but require time-consuming calculations for key parameters such as sound velocity and the Grüneisen parameter. This work presents a high-throughput framework, physical-informed kappa (PINK), which combines the predictive power of crystal graph convolutional neural networks (CGCNNs) with the physical interpretability of the Slack model to predict $\kappa_L$ directly from crystallographic information files (CIFs). Unlike previous approaches, PINK enables rapid, batch predictions by extracting material properties such as bulk and shear modulus from CIFs using a well-trained CGCNN model. These properties are then used to compute the necessary parameters for $\kappa_L$ calculation through a simplified physical formula. PINK was applied to a dataset of 377,221 stable materials, enabling the e icient identification of promising candidates with ultralow $\kappa_L$ values, such as $Ag_3Te_4W$ and $Ag_3Te_4Ta$. The platform, accessible via a user-friendly interface, offers an unprecedented combination of speed, accuracy, and scalability, significantly accelerating material discovery for thermal management and energy conversion applications.

## INTRODUCTION

Understanding the temperature dependence of lattice thermal conductivity ($\kappa_L$) is essential for assessing the thermal transport capabilities of a material. This property plays a crucial role in both scientific research and industrial applications, including thermal management in microelectronics[1,2], energy conversion[3], and temperature regulation[4]. For example, materials exhibiting high $\kappa_L$, such as boron arsenide (BAs), are particularly suitable for heat dissipation in gallium nitride devices[5]. Conversely, materials with low $\kappa_L$ can improve thermoelectric conversion efficiency by enabling the effective transformation of waste heat into electrical energy[6].

In recent years, significant theoretical advancements have been made in the theoretical prediction of $\kappa_L$ in solid materials[7-10]. A widely used approach for predicting $\kappa_L$ involves solving the phonon Boltzmann transport equation (PBTE) within the framework of density functional theory (DFT)[9], while classical molecular dynamics (MD) simulations are particularly useful for systems with complex crystal structures[11]. However, identifying materials with exceptionally low or high $\kappa_L$ remains a significant challenge, mainly due to the high computational costs and time-consuming synthesis processes[12]. Moreover, calculations required to obtain interatomic force constants (IFCs) are especially demanding for large, low-symmetry primitive cells[13]. Furthermore, the reliability of MD simulations is strongly dependent on the selection of interatomic potentials, limiting their broader applicability[14]. Besides these challenges, significant progress has been made in accelerating material discovery and improving performance. Luo *et al.* reviewed the application of machine learning (ML) for predicting $\kappa_L$, emphasizing the potential of high-throughput predictions and ML potentials (MLPs) to overcome the limitations of traditional approaches[13]. Liu *et al.* focused on active and reversible techniques for regulating $\kappa_L$, such as the use of ferroelectric, ferromagnetic, and nanomaterials, enabling dynamic control of thermal conductivity for efficient thermal management[15]. Additionally, Shi *et al.* examined advancements in thermoelectric materials for multifunctional energy conversion and storage technologies, highlighting ongoing challenges related to scalability, material stability, and efficiency that must be addressed to fully realize their potential in practical applications[16]. Consequently, rapid determination of $\kappa_L$ is crucial for advancing these materials.

Alternatively, empirical models such as the Debye-Callaway model[17,18] and the Slack model[19,20] provide faster and more cost-effective approaches for estimating $\kappa_L$. The Slack model, in particular, has been widely applied to predict $\kappa_L$ in a variety of materials[21-26]. For instance, Qin *et al.* successfully employed the model to quickly predict thermal conductivity, offering valuable insights into thermal transport behavior[27]. Cao *et al.* explored the n-type thermoelectric properties of $ABO_3$ cubic chalcogenides using a high-throughput method combined with Slack modeling[28]. They screened 46 stable materials, identified four conduction band minima structures, investigated the influence of chemical bonding on transport properties, and shortlisted 13 candidates with high thermoelectric figure of merit (ZT) values. However, the model's reliance on experimental data or first-principles calculations for several parameters limits its scalability for large-scale, high-throughput screenings. Obtaining critical parameters, such as average sound velocity, acoustic Debye temperature, and the Grüneisen parameter, often requires considerable time and resources, posing a significant barrier.

In our previous work[29], we proposed a refined formula based on the Slack model, which enables highly accurate predictions of $\kappa_L$ with an 8.97% mean relative error. The formula utilizes only the shear modulus, average sound velocity, and Grüneisen parameter, all of which are relatively easy to obtain. For example, the bulk modulus ($B$) and shear modulus ($G$) can be used to derive the average sound velocity in a material[30]. Additionally, significant research has been conducted to simplify the estimation of the Grüneisen parameter. Belomestnykh[31] developed a method that links Poisson's ratio with sound velocity and elastic properties, yielding results consistent with quasiharmonic lattice dynamics calculations. This work underscores the strong relationship between elastic modulus and $\kappa_L$[29]. The crystallographic information file (CIF) provides comprehensive data on crystal structures, including lattice constants, crystal systems, density, and other key parameters. However, existing approximation methods have not fully exploited this information to predict $\kappa_L$. Only a limited number of studies have directly connected CIF data with $\kappa_L$ for fast, high-throughput predictions. For example, Ju *et al.* used a neural network that leveraged descriptors from a pre-trained model to establish a relationship between crystal information and thermal conductivity[32]. Xie *et al.* introduced the crystal graph convolutional neural network (CGCNN) method[33], which converts crystal structure data into graph representations, enabling convolutional neural networks to predict the relationship between crystal features and thermoelectric properties[34,35]. Recently, Omee *et al.* reviewed the performance of five out-of-distribution (OOD) test sets across eight graph neural network (GNN) models using elasticity datasets[36]. Notably, the CGCNN model achieved the best mean absolute error (MAE) for both the Leave-One-Cluster-Out (LOCO[37]) test [0.0585 log10 (GPa)] and the SparseXsingle test [0.0499 log10 (GPa)], targeting structures with the lowest density and surpassing the other seven GNN models. These findings highlight the excellent generalization capabilities of CGCNN models.

In this work, to enable rapid and high-throughput $\kappa_L$ predictions, we integrate the physical interpretability of our derived formula with the predictive power of the CGCNN model. This study introduces a high-throughput framework that combines a trained modulus model with our formula, facilitating the fast estimation of $\kappa_L$ directly from CIF files. Encapsulated in a custom-developed web application, physical-informed kappa (PINK), this process enables batch predictions of $\kappa_L$ within seconds of uploading CIF files. Users can also customize inputs such as bulk modulus, shear modulus, and Grüneisen parameter.

Our framework begins by extracting crystallographic information from the CIF files and utilizing the trained CGCNN model to predict the bulk and shear modulus. Subsequently, physical models are applied to calculate the average sound velocity and Grüneisen parameters, which are then incorporated into a formula to calculate the $\kappa_L$. Using this approach, we predict $\kappa_L$ for 377,221 stable materials identified by Merchant *et al.* through graph networks[38]. Building on these high-throughput predictions, we develop an efficient method to accelerate the screening of materials with ultralow $\kappa_L$, applicable to any inorganic crystal structure with one or more CIF files. This method enabled the identification of thousands of promising materials with low $\kappa_L$ from over 370,000 inorganic crystalline samples, with minimal computational cost. To validate our results, we confirm the ultralow $\kappa_L$ values for $Ag_3Te_4X$ (X = W, Ta) through first-principles calculations. The PINK application, powered by the CGCNN model, serves as a powerful tool for rapid material pre-screening. It provides researchers with an efficient, user-friendly platform for estimating $\kappa_L$, accelerating the discovery of materials with optimal thermal properties.

## MATERIALS AND METHODS

### CGCNN algorithms

Before presenting the framework of PINK, it is essential to clarify the method by which the CGCNN model predicts material properties based on crystal structures. CGCNN, an advanced ML algorithm, uses trained models to predict material properties with high efficiency[33]. The crystal structure is converted into a graph representation, where nodes correspond to atoms, and edges represent the bonds between them. This format allows the model to capture the local chemical environment.

Through convolutional and pooling layers, CGCNN autonomously identifies critical features necessary for predicting various material properties, such as bulk modulus and shear modulus. These predictions are both accurate and interpretable, providing valuable insights for the rational design of new materials. Moreover, the robust generalization capabilities of this model enable it to handle diverse crystal structures and compositions, significantly accelerating the material discovery process[33,34,39]. In this study, the elastic modulus dataset was split into training, validation, and test sets with a ratio of 80%, 10%, and 10%, respectively. The model consisted of three convolutional layers and two hidden layers, and was trained for 30 epochs with a learning rate of 0.01.

### The PINK framework

As illustrated in Figure 1, we present a comprehensive workflow for calculating $\kappa_L$ using our automated property prediction system. This workflow is designed to be user-friendly, requiring only CIF files as input. To ensure accurate calculations, the system automatically converts the uploaded crystal structure into its primitive cell format, which is essential for both CGCNN predictions and the parameters used in Equation (2). The process begins by extracting fundamental crystallographic data from the CIF file, including the primitive cell volume, number of atoms, and density. Next, our embedded CGCNN model, trained on extensive material data, predicts the bulk modulus and shear modulus. Using these predicted values, custom Python scripts calculate additional physical parameters crucial for estimating $\kappa_L$, including the longitudinal and transverse sound velocities, the average speed of sound, and the Grüneisen parameters. Finally, all of these calculated quantities are systematically incorporated into Equation (2) to compute $\kappa_L$. This automated workflow significantly streamlines the process of $\kappa_L$ prediction, making it accessible to researchers without requiring in-depth expertise in each individual computational step.

The application provides comprehensive physical property data for 377,221 new materials, including 11,869 materials screened in this study. The modified open-source CGCNN code used for predicting bulk and shear modulus, as well as the Python scripts for CIF file processing and calculation execution (e.g., "app.py"), is also available. All of these data and codes are accessible via the following link: https://github.com/Jack-Liu0227/AI4Kappa.

### Surrogate an interpretable formula for $\kappa_L$

Recently, Wang *et al.* proposed a simple and universal empirical formula that exhibits strong generalization ability and provides clear physical insights for $\kappa_L$ of crystals, which is given as[29]:

$$\kappa_L = \frac{G\upsilon_s V^{\frac{1}{3}}}{nT^{\delta}} \cdot e^{-\gamma}, \tag{1}$$

where $G$ is the shear modulus, $\upsilon_s$ represents the average sound velocity, $V$ is the volume of the primitive cell, $n$ is the number of atoms in the primitive cell, δ lies between 1 and 2 (with δ = 1 for three-phonon scattering), $T$ is the temperature in Kelvin, and $\gamma$ denotes the Grüneisen parameter. It is important to note that $\kappa_L$ and $\upsilon_2$ in Equation (1) do not exhibit a conventional proportional correlation, as both $G$ and $\gamma$ are functionally dependent on $\upsilon$ (see Supplementary Materials for details)[29].

The theoretical basis of the power law is complex, involving competition between scattering processes driven by cubic and quartic anharmonic terms[40,41]. For simplicity, we focus only on three-phonon scattering, assuming δ = 1:

$$\kappa_L = \frac{G\upsilon_s V^{\frac{1}{3}}}{nT} \cdot e^{-\gamma}, \tag{2}$$

Which, derived from Slack's approach[23] is useful for evaluating $\kappa_L$ across various materials. A key aspect in evaluating $\kappa_L$ involves determining the average speed of sound ($\upsilon_s$) and the Grüneisen parameter ($\gamma$). Jia *et al.* proposed that $\upsilon_s$ can be accurately estimated from elastic properties [bulk modulus ($B$) and shear modulus ($G$)][42]. This approach is computationally more efficient than experimental methods or costly lattice dynamics simulations. The bulk modulus ($B$) and shear modulus ($G$) can both be extracted from our trained CGCNN model, providing an alternative means of estimating elastic properties and sound velocities, as demonstrated in[27,42].

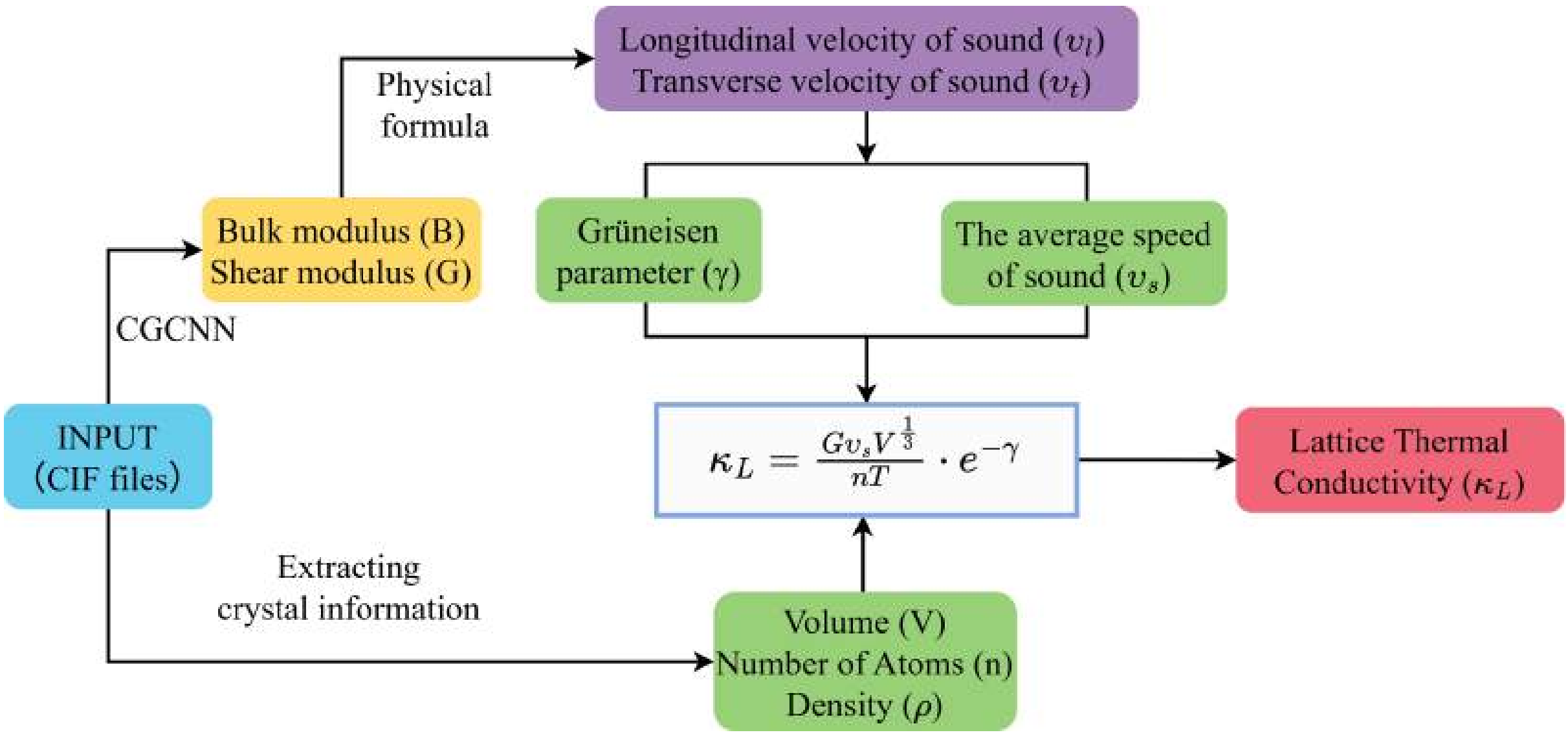

**Figure 1.** The workflow for calculating $\kappa_L$ using PINK begins with the input of CIF files representing crystal structures. Starting with these CIF files, the framework utilizes CGCNN to predict the bulk and shear modulus, while also extracting crystal information such as volume, number of atoms, and density. These parameters are subsequently used to calculate both longitudinal and transverse sound velocities, which are essential for determining the Grüneisen parameter and the average speed of sound. All of these parameters are incorporated into Equation (2), which includes the Grüneisen parameter ($\gamma$), volume ($V$), temperature ($T$), and other variables necessary for predicting $\kappa_L$. PINK: Physical-informed kappa; CIF: crystallographic information file; CGCNN: crystal graph convolutional neural network.

$$v_l = \sqrt{\frac{B + \frac{4}{3}G}{\rho}}, \tag{3}$$

$$v_t = \sqrt{\frac{G}{\rho}}, \tag{4}$$

$$v_s = \left\{\frac{1}{3}\left[\frac{1}{v_l^3} + \frac{2}{v_t^3}\right]\right\}^{-\frac{1}{3}}, \tag{5}$$

where $v_s$, $v_l$, and $v_t$ are the average sound velocity, longitudinal sound velocity, and transverse sound velocity, respectively, and $\rho$ is the material density.

After estimating $v_s$ from the bulk modulus ($B$) and shear modulus ($G$), the next step is to determine the Grüneisen parameter, which quantifies the anharmonicity of the material[43]. The speed of sound serves as an indicator of the strength of atomic interactions, with weaker interactions generally leading to lower sound velocities. It has been shown that the relationship between Poisson's ratio ($\nu$) and $\gamma$ is as follows[31,44]:

$$\nu = \frac{x^2 - 2}{2x^2 - 2}, \quad \gamma = \frac{3}{2}\left(\frac{1+\nu}{2-3\nu}\right). \tag{6}$$

where $x$ represents the ratio of longitudinal to transverse sound velocity, $x = v_l/v_t$.

Using the above method, both $v_s$ and $\gamma$ can be estimated quickly from elastic properties, particularly the shear and bulk modulus. Previous studies indicate that this approach aligns well with experimental results for cubic, isotropic, and quasi-isotropic structures[42,45,46].

### Deployment of PINK for $\kappa_L$

Streamlit is an open-source Python library that simplifies the creation of custom web apps for data-driven applications. It facilitates rapid development of interactive apps by converting Python scripts into shareable web applications in just a few minutes. Figure 2 illustrates the process of deploying an app: first, upload the project code to GitHub, and then create the application on the Streamlit platform by selecting the relevant project branch and Python file (e.g., app.py) to run. Setting up the application environment can be challenging, as web applications typically require multiple Python packages with specific versions. Fortunately, our code streamlines this process by including a requirements.txt file to ensure all dependencies are installed correctly. This allows the application to be deployed entirely in Python without requiring front-end experience. By leveraging Equation (2) and the CGCNN model, we developed the $\kappa_L$ calculation application based on this framework.

After deploying the application, users can quickly calculate a material's elastic properties, $\kappa_L$, and other relevant outputs by uploading CIF files. The results are displayed on the website in a DataFrame format, and users can download them as CSV files. An illustration of the program's interface is shown in Figure 3. Importantly, the app supports uploading single or multiple CIF files simultaneously, running the entire framework in parallel to provide results for all materials at once.

Our application, PINK, is easily accessible via the following link: https://kappap-ai.streamlit.app, which can be used both on your phone and on your local computer. If you prefer to deploy it locally or on Streamlit's server, please refer to the README.md for detailed instructions on setting up the software.

### Calculating $\kappa_L$ using *ab initio* study

We implemented *ab initio* study through the Vienna Ab-initio Simulation Package (VASP)[47]. The calculations incorporated the projector augmented-wave (PAW) approach combined with the Perdew-Burke-Ernzerhof (PBE) functional for exchange-correlation[48-50]. To achieve high computational precision, we selected a 520 eV planewave cutoff energy alongside a Monkhorst-Pack sampling with a 4 × 4 × 4 $k$-mesh. The computational parameters were optimized with convergence thresholds of $10^{-8}$ eV for total energy and $10^{-4}$ eV/Å for atomic forces. In determining the second-order IFCs, our calculations utilized a supercell configuration of 2 × 2 × 2, employing the finite displacement methodology with a 4 × 4 × 4 $k$-point mesh. The third-order software package[40] was subsequently used to extract the third-order IFCs. The $\kappa_L$ calculations, which account for three-phonon scattering processes, were performed using ShengBTE with a dense 20 × 20 × 20 $q$-point sampling[51].

## RESULTS AND DISCUSSION

To streamline the time-intensive process of learning CGCNN for predicting material properties and handling file processing, we developed a high-throughput framework encapsulated in a user-friendly application. The interface allows researchers to input single or multiple CIF files simultaneously, generating instantaneous $\kappa_L$ predictions for the specified compounds. The efficiency of this framework arises from its integration of pre-trained CGCNN models with Equation (2), providing rapid assessments of thermal transport properties.

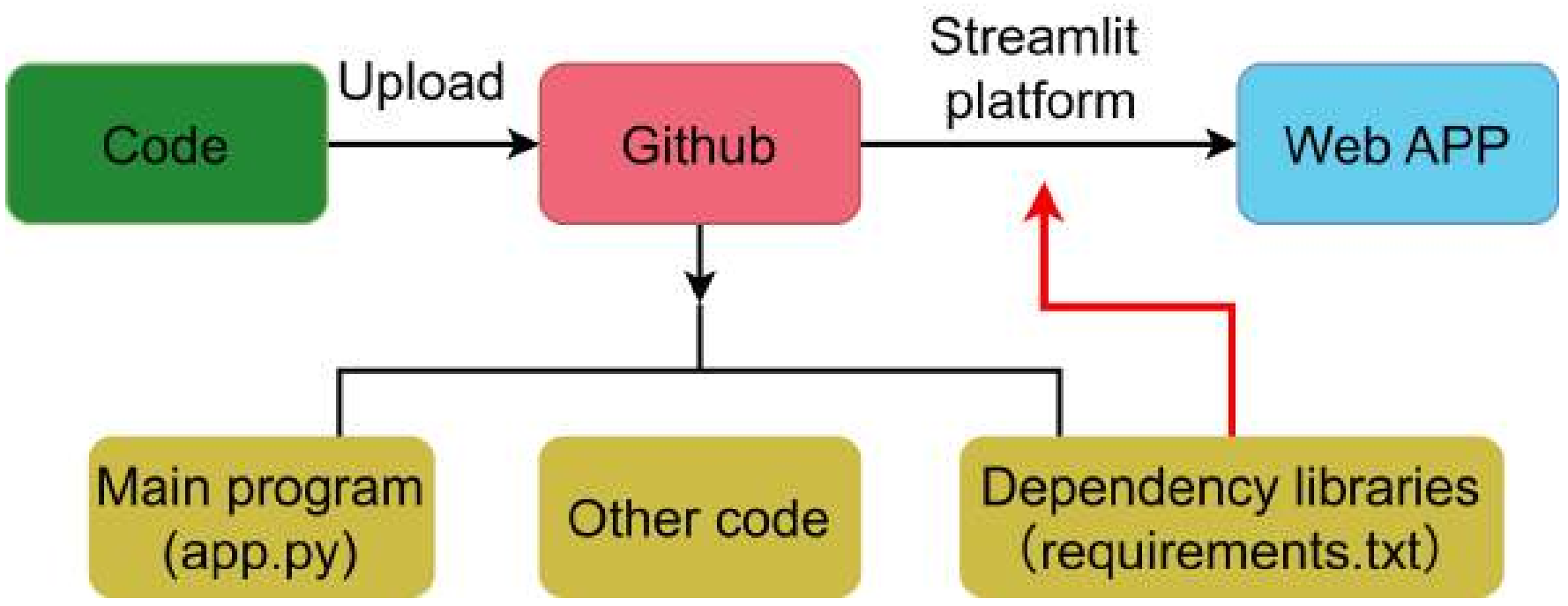

**Figure 2.** PINK code deployment process. To deploy and run the web application, one first uploads the code - along with the “app.py”, “requirements.txt”, and any other necessary files to GitHub. Then, we use the Streamlit platform to deploy the application online. PINK: Physical-informed kappa.

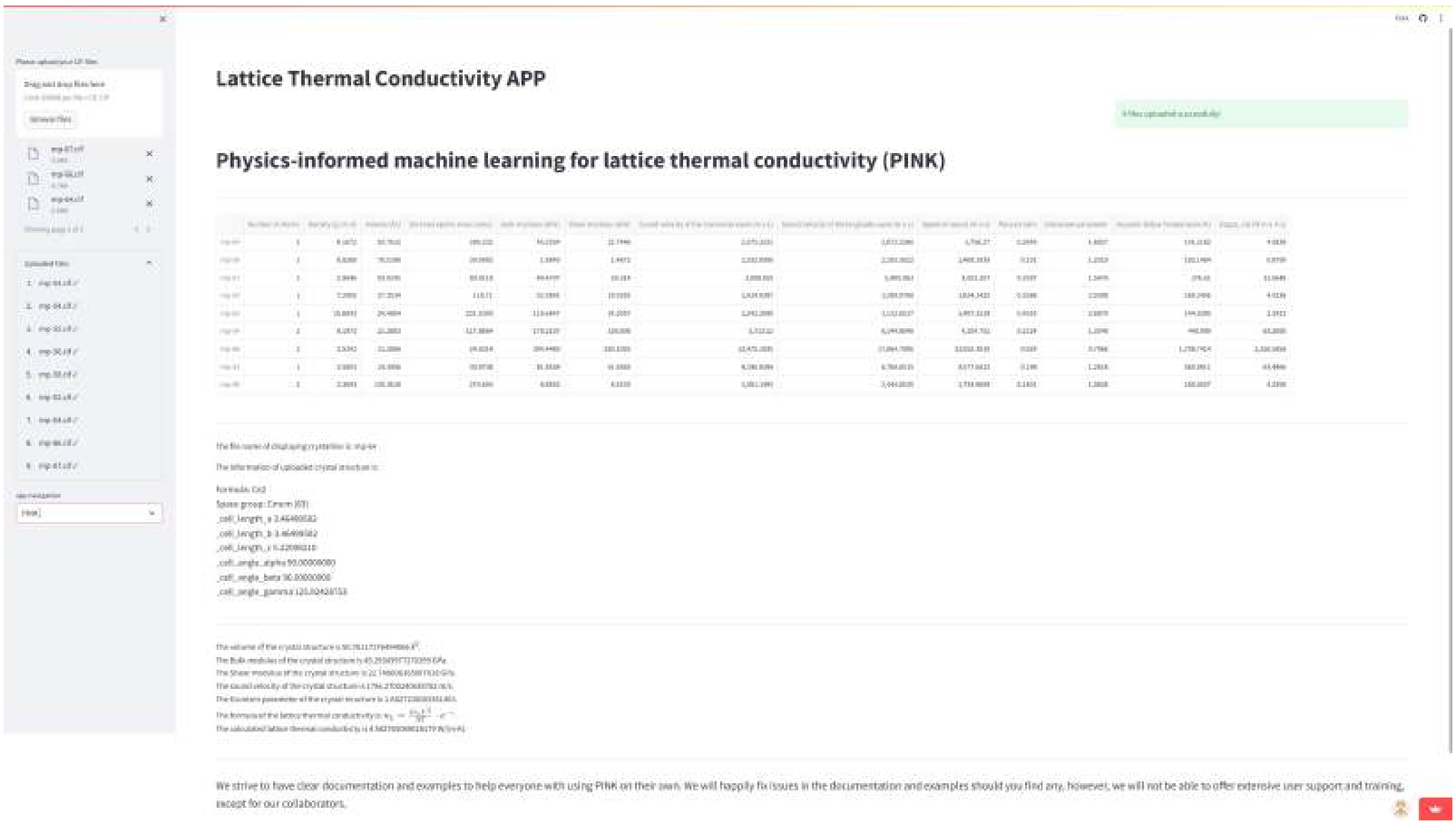

**Figure 3.** The web page for our PINK app is divided into two panels. The left panel allows users to upload files, while the right panel displays the results. The output includes a DataFrame that lists various properties such as the number of atoms, density (g/cm$^{-3}$), volume (Å$^3$), atomic mass (amu), bulk modulus (GPa), shear modulus (GPa), transverse and longitudinal wave sound velocities (m/s), speed of sound (m/s), Poisson’s ratio ($v$), Grüneisen parameter ($\gamma$), acoustic Debye temperature ($\theta_a$, K), and lattice thermal conductivity (W·m$^{-1}$·K$^{-1}$). For detailed instructions on using PINK, please refer to the PINK_tutorial.mp4. Additionally, the app supports custom functions for calculating bulk modulus (GPa), shear modulus (GPa), and Grüneisen parameter, with a separate tutorial available in PINK_Custom_Parameters_tutorial.mp4. PINK: Physical-informed kappa.

Given that thermoelectric performance is strongly influenced by materials with low $\kappa_L$, we conducted a high-throughput screening across material dataset. This systematic evaluation successfully identified 11,869 potential candidates with promising thermal transport characteristics. To validate our screening approach, we selected $Ag_3Te_4X$ (X = W, Ta) from a specific ternary system and performed detailed *ab initio* calculations to verify their properties.

### Data collection

For evaluating ML performance in materials science applications, we utilized the Matbench V0.1[52]. Our analysis focused on two specific datasets within this collection that address elastic properties: "matbench_log_gvrh" and "matbench_log_kvrh". These elastic modulus datasets contain identical material entries (10,987 in total) and are specifically designed to predict the logarithmic values of shear modulus ($G$) and bulk modulus ($B$) using the Voigt-Reuss-Hill (VRH) averaging methodology. The comprehensive nature of these standardized collections makes them ideal for training our ML models to predict key elastic characteristics. A detailed analysis of the dataset is presented in Figure 4, which illustrates the statistical distribution across three key aspects: the classification of crystal systems, the number of atoms in the primitive cell, and the distribution of chemical elements. The dataset exhibits remarkable diversity, incorporating materials from all seven fundamental crystal systems and 84 different elements in various structural arrangements.

For predicting $\kappa_L$, we utilized datasets obtained from the AFLOW database[53] and relevant publications[20,54-57], which include both experimentally measured values and computationally derived properties. The AFLOW database provides comprehensive information on crystal structures and thermal characteristics, with $\kappa_L$ values calculated using the methodologies outlined in[29,58]. To construct a test dataset for evaluating our application, we collected crystal structures, Grüneisen parameters ($\gamma$), and their corresponding $\kappa_L$ values from AFLOW and other literature.

Obtaining high-throughput datasets for computational materials science can be challenging. However, recent advancements in ML have significantly enhanced the discovery of stable materials. Merchant *et al.* employed deep learning and GNNs to scale materials discovery, particularly for inorganic crystals[38]. Their work expanded the known set of stable materials by adding 381,000 new entries to the convex hull, resulting in a total of 377,221 stable crystal structures - a tenfold increase over previous datasets.

We accessed their extensive dataset through GitHub: https://github.com/google-deepmind/materials_discovery. The repository includes 377,221 valid CIF files in the "by_composition" folder, compatible with CGCNN, and a summary CSV file containing bandgap, crystal symmetry, and decomposition energy data. These materials encompass compositions ranging from two to six elements, with atomic numbers spanning from 2 to 106. Additional data from the Materials Project further complements these datasets, enabling the targeted retrieval of material properties through its open-source API. Together, these resources provide a robust foundation for high-throughput computations and analyses.

### Model evaluation of CGCNN

The interpretable formula, detailed in the methods section, elucidates the correlation between elastic modulus and $\kappa_L$. In ML, MAE and $R^2$ (R-squared) are standard metrics for evaluating regression models. MAE quantifies the average magnitude of prediction errors and is defined as:

$$\mathrm{MAE} = \frac{1}{n}\sum_{i=1}^{n} |y_i - \hat{y}_i|, \tag{7}$$

where $y_i$ denotes the actual values and $\hat{y}_i$ the predicted values. Lower MAE values denote higher prediction accuracy. $R^2$ assesses the proportion of variance in the dependent variable explained by:

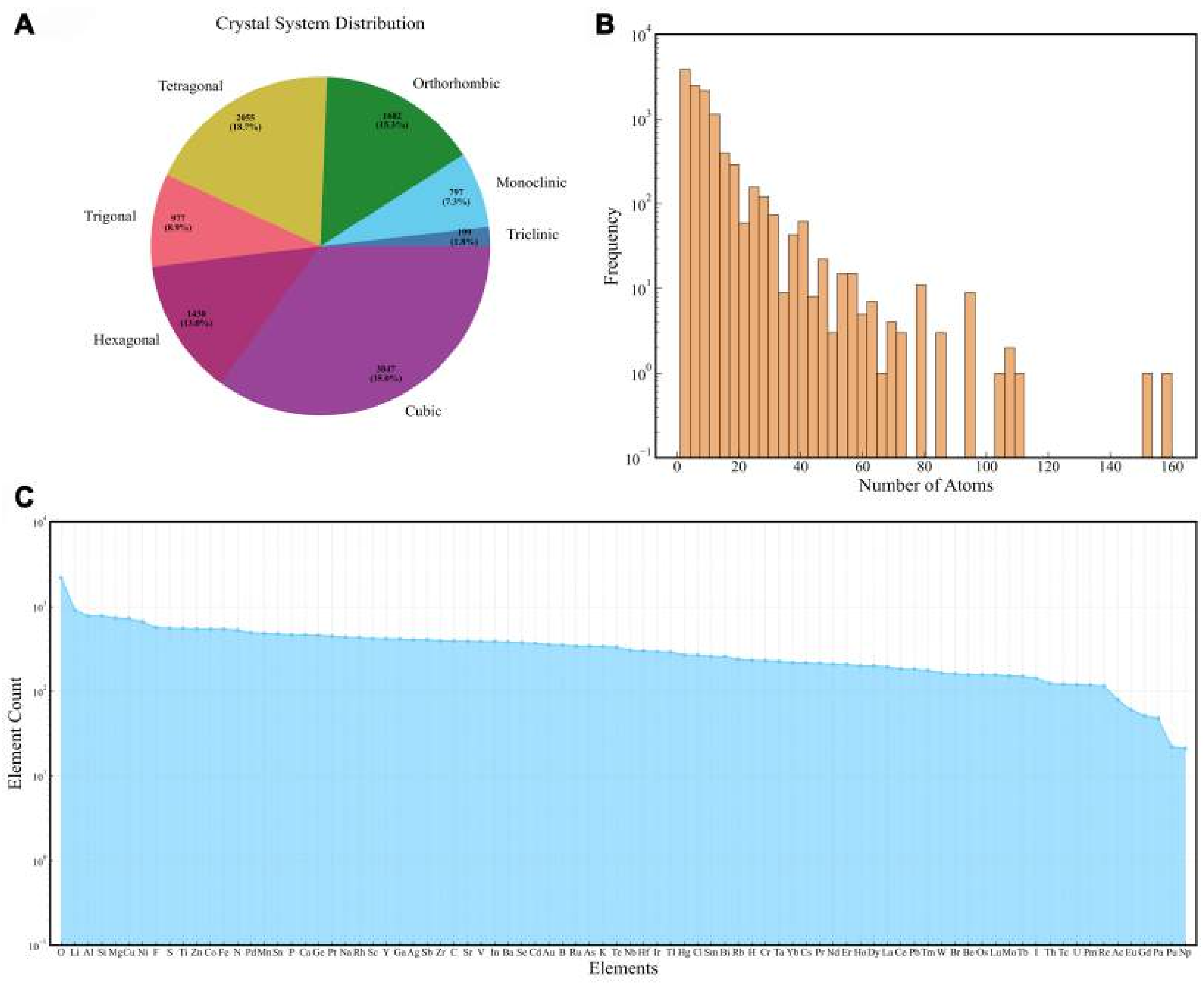

**Figure 4.** Statistical analysis of the training dataset. (A) The distribution of seven crystal systems, with cubic being the most common (3,847 structures), followed by tetragonal (2,055 structures), while triclinic is the least one (199 structures); (B) Distribution of range of number of atoms in the primitive cell (1-160 atoms) across the dataset; (C) Elemental distribution that illustrates the frequency of 84 distinct elements. The dataset encompasses transition metals, main group elements, and rare earth elements, with oxygen showing the highest frequency.

$$R^2 = 1 - \frac{\sum_{i=1}^{n}(y_i - \hat{y}_i)^2}{\sum_{i=1}^{n}(y_i - \bar{y})^2}. \tag{8}$$

where $\bar{y}$ represents the mean of the actual values. $R^2$ values approaching 1 indicate a better model fit.

The performance of the embedded CGCNN model on the test dataset is illustrated in Figure 5. The MAE for both the shear and bulk moduli is below 13, with $R^2$ values approaching 1, indicating a strong correlation between the predicted and DFT-calculated elastic modulus. These results demonstrate the model's reliability and predictive accuracy.

### Model evaluation of $\kappa_L$

After predicting the shear and bulk moduli using the trained CGCNN model, the approximate average speed of sound was estimated. Utilizing known crystal structure information, we applied Equation (2) to approximate the material's $\kappa_L$. To validate the PINK application, we compared its predictions with $\kappa_L$ values calculated via DFT for 2,535 materials from the AFLOW database[53] and 46 experimentally measured values from the literature[20,54-57]. The Grüneisen parameters were obtained from the AFLOW database and experimental data, respectively.

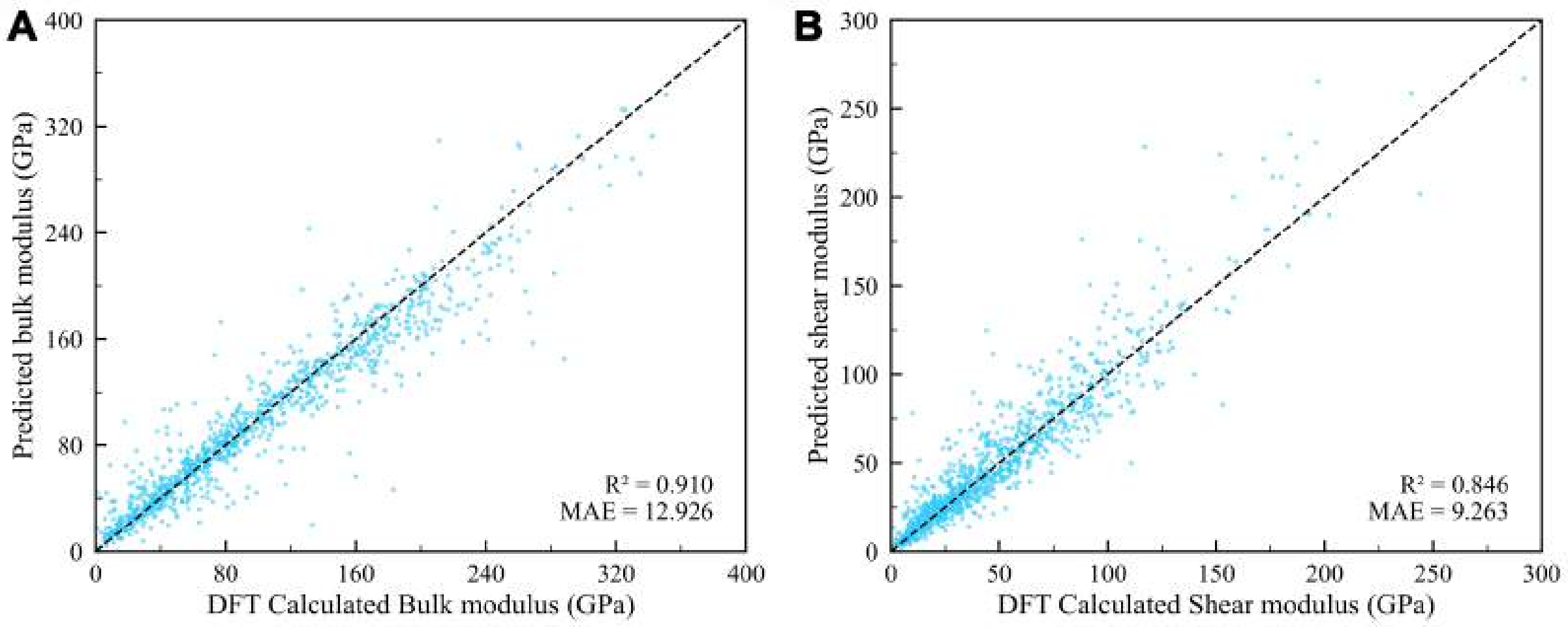

**Figure 5.** The comparison of predicted *vs.* DFT-calculated values for (A) bulk modulus and (B) shear modulus across the test dataset of 10,987 structures. DFT: Density functional theory.

Figure 6 presents scatter plots comparing $\kappa_L$ predictions by PINK with calculated and experimentally measured values. In Figure 6A, each point represents a material, with the solid diagonal line indicating perfect agreement between predicted and calculated values. The dashed lines denote an acceptable range of deviation. Within the dataset, 2,415 points (95.27%) fall within this range, highlighting the model's high accuracy. The clustering of points near the diagonal line further confirms a strong correlation between PINK predictions and DFT calculations. Deviations are likely attributable to the inapplicability of certain materials to the Slack model or inaccuracies in the elastic modulus predictions[43].

In Figure 6B, $\kappa_L$ predictions from PINK are compared with experimental values, with each point labeled by the corresponding material. Similar to Figure 6A, the solid diagonal line represents perfect agreement, and the dashed lines denote acceptable deviation boundaries. The model achieves a MAE of 0.526 and an $R^2$ value of 0.881, indicating a close correspondence between PINK predictions and experimental results. The clustering of data points near the diagonal line demonstrates that PINK effectively predicts $\kappa_L$ across diverse materials and crystal symmetries.

For additional validation, Table 1 presents the predicted $\kappa_L$ values alongside their experimental counterparts for 46 materials. These results further substantiate the reliability and effectiveness of PINK in predicting $\kappa_L$, showing close alignment with both DFT-calculated and experimentally measured values. Notably, the accuracy of $\kappa_L$ predictions could be significantly enhanced with more precise Grüneisen parameters and improved predictions of shear and bulk moduli[29].

### Comparison of calculation time for $\kappa_L$

To demonstrate $\kappa_L$ the efficiency of prediction application, PINK, we compared its computational time against other commonly used methods. Traditional approaches, such as solving the PBTE with second- and third-order force constants[60] or employing the equilibrium MD Green-Kubo, typically require several hours for simple systems and days to weeks for complex ones.

**Table 1. Predicted and experimental room-temperature $\kappa_L$ values for compounds from the literature[20,54-57] are presented**

| Materials | ID-number | $n$ | $\rho$ (g·cm$^3$) | $V$ (Å$^3$) | $G$ (GPa) | $\upsilon_s$ (m·s$^{-1}$) | $\gamma$ | $\kappa_{exp}$ (W·m$^{-1}$·K$^{-1}$) | $\kappa_{PINK}$ (W·m$^{-1}$·K$^{-1}$) |
|---|---|---|---|---|---|---|---|---|---|
| AgCl[54] | mp-22922 | 2 | 5.583 | 42.631 | 8.801 | 1,423.597 | 1.900 | 1 | 1.091 |
| AlAs[20] | mp-2172 | 2 | 3.591 | 47.126 | 40.510 | 3,733.087 | 0.660 | 98 | 47.054 |
| AlSb[20] | mp-2624 | 2 | 4.078 | 60.561 | 28.808 | 2,959.047 | 0.600 | 56 | 30.620 |
| BN[20] | mp-1639 | 2 | 3.458 | 11.919 | 350.203 | 10,995.529 | 0.700 | 760 | 727.981 |
| BP[20] | mp-1479 | 2 | 2.953 | 23.500 | 201.292 | 9,011.073 | 0.923 [by Equation (6)] | 350 | 344.207 |
| C[20] | mp-66 | 2 | 3.496 | 11.410 | 547.436 | 13,613.398 | 0.750 | 3,000 | 1,320.871 |
| CaO[20] | mp-2605 | 2 | 3.287 | 28.332 | 63.316 | 4,863.622 | 1.570 | 27 | 32.552 |
| CdTe[20] | mp-406 | 2 | 5.473 | 72.827 | 14.349 | 1,818.434 | 0.520 | 7.5 | 10.797 |
| GaAs[20] | mp-2534 | 2 | 5.053 | 47.532 | 45.152 | 3,291.363 | 0.750 | 45 | 42.381 |
| GaP[20] | mp-3490 | 2 | 4.006 | 41.737 | 48.296 | 3,846.005 | 0.750 | 100 | 50.725 |
| GaSb[20] | mp-1156 | 2 | 5.288 | 60.133 | 28.033 | 2,557.748 | 0.750 | 40 | 22.115 |
| Ge[20] | mp-32 | 2 | 5.042 | 47.847 | 47.134 | 3,362.109 | 1.060 | 65 | 33.219 |
| InAs[20] | mp-20305 | 2 | 5.336 | 59.050 | 23.662 | 2,355.122 | 0.570 | 30 | 20.454 |
| InP[20] | mp-20351 | 2 | 4.582 | 52.840 | 27.554 | 2,742.762 | 0.600 | 93 | 25.940 |
| InSb[20] | mp-20012 | 2 | 5.384 | 72.965 | 17.668 | 2,026.668 | 0.560 | 20 | 14.244 |
| KBr[20] | mp-23251 | 2 | 2.624 | 75.294 | 6.156 | 1,709.446 | 1.450 | 3.4 | 1.737 |
| KBr[20] | mp-570891 | 2 | 2.989 | 66.111 | 8.391 | 1,885.943 | 1.450 | 3.4 | 2.502 |
| KCl[20] | mp-23193 | 2 | 1.904 | 65.033 | 6.387 | 2,050.803 | 1.450 | 7.1 | 2.059 |
| KI[20] | mp-22898 | 2 | 2.972 | 92.743 | 5.790 | 1,546.914 | 1.450 | 2.6 | 1.585 |
| LiF[20] | mp-1009009 | 2 | 2.569 | 16.768 | 58.174 | 5,236.933 | 1.500 | 17.6 | 28.999 |
| LiH[20] | mp-23703 | 2 | 0.825 | 16.002 | 39.346 | 7,537.234 | 1.280 | 15 | 34.630 |
| MgO[20] | mp-1265 | 2 | 3.471 | 19.279 | 123.811 | 6,564.689 | 1.440 | 60 | 86.060 |
| NaBr[20] | mp-22916 | 2 | 3.121 | 54.749 | 14.495 | 2,392.521 | 1.500 | 2.8 | 4.897 |
| NaCl[20] | mp-22862 | 2 | 2.105 | 46.096 | 16.650 | 3,123.706 | 1.560 | 7.1 | 6.531 |
| NaF[20] | mp-682 | 2 | 2.693 | 25.894 | 21.929 | 3,171.392 | 1.500 | 18.4 | 7.651 |
| NaI[20] | mp-23268 | 2 | 3.572 | 69.675 | 6.823 | 1,547.000 | 1.560 | 1.8 | 1.521 |
| PbS[20] | mp-21276 | 2 | 7.334 | 54.174 | 26.485 | 2,111.102 | 2.000 | 2.9 | 4.772 |
| PbSe[20] | mp-2201 | 2 | 7.886 | 60.254 | 22.618 | 1,876.743 | 1.500 | 2 | 6.189 |
| RbBr[20] | mp-22867 | 2 | 3.164 | 86.781 | 6.905 | 1,651.509 | 1.450 | 3.8 | 1.974 |
| RbCl[20] | mp-23295 | 2 | 2.672 | 75.148 | 7.337 | 1,854.222 | 1.450 | 2.8 | 2.244 |
| RbI[20] | mp-22903 | 2 | 3.360 | 104.957 | 6.228 | 1,517.862 | 1.410 | 2.3 | 1.814 |
| Si[20] | mp-149 | 2 | 2.281 | 40.888 | 55.828 | 5,480.852 | 1.060 | 166 | 60.869 |
| SiC[20] | mp-8062 | 2 | 3.227 | 20.635 | 222.879 | 9107.212 | 0.750 | 490 | 438.312 |
| SrO[20] | mp-2472 | 2 | 4.878 | 35.277 | 47.864 | 3,468.152 | 1.520 | 12 | 19.845 |
| TePb[20] | mp-19717 | 2 | 7.857 | 70.758 | 17.511 | 1,674.996 | 2.009 [by Equation (6)] | 2.5 | 2.711 |
| ZnS[20] | mp-10695 | 2 | 3.999 | 40.476 | 32.768 | 3,191.504 | 0.750 | 27 | 28.268 |
| ZnSe[20] | mp-1190 | 2 | 5.064 | 47.338 | 31.048 | 2,763.063 | 0.750 | 19 | 24.432 |
| ZnTe[20] | mp-2176 | 2 | 5.419 | 59.146 | 25.383 | 2,416.126 | 0.970 | 18 | 15.097 |
| AlN[20] | mp-661 | 4 | 3.201 | 42.527 | 135.549 | 7,197.895 | 0.700 | 350 | 140.931 |
| BeO[20] | mp-2542 | 4 | 2.967 | 27.992 | 123.309 | 7,116.682 | 0.750 | 370 | 104.885 |
| CdS[20] | mp-672 | 4 | 4.576 | 104.863 | 16.885 | 2,166.455 | 0.750 | 16 | 6.790 |
| GaN[20] | mp-804 | 4 | 5.924 | 46.943 | 110.843 | 4,794.507 | 0.700 | 210 | 79.333 |
| ZnO[20] | mp-2133 | 4 | 5.438 | 49.719 | 33.376 | 2,790.196 | 0.750 | 60 | 13.479 |
| $Bi_2Te_3$[55] | mp-34202 | 5 | 7.315 | 181.767 | 10.489 | 1,339.685 | 1.490 | 1.6 | 1.196 |
| $Al_2O_3$[56] | mp-1143 | 10 | 3.873 | 87.420 | 132.798 | 6,501.426 | 1.340 | 30 | 33.445 |
| ZnSb[57] | mp-753 | 16 | 6.347 | 391.759 | 32.387 | 2,518.332 | 1.681 [by Equation (6)] | 3.5 | 2.315 |

The CIFs for these materials were obtained from the Materials Project. Here, $n, \rho, V$ and $G$ are the number of atoms in the primitive cell, the density, the volume of the primitive cell and the shear modulus predicted by CGCNN, respectively. $\upsilon_s$ represents the average speed of sound calculated from Equation (5), $\gamma$ is the experimental Grüneisen parameter, $\kappa_{exp}$ denotes the experimentally measured values, and $\kappa_{PINK}$ is the calculated $\kappa_L$ from Equation (2). PINK: Physical-informed kappa; CIFs: crystallographic information file; CGCNN: crystal graph convolutional neural network.

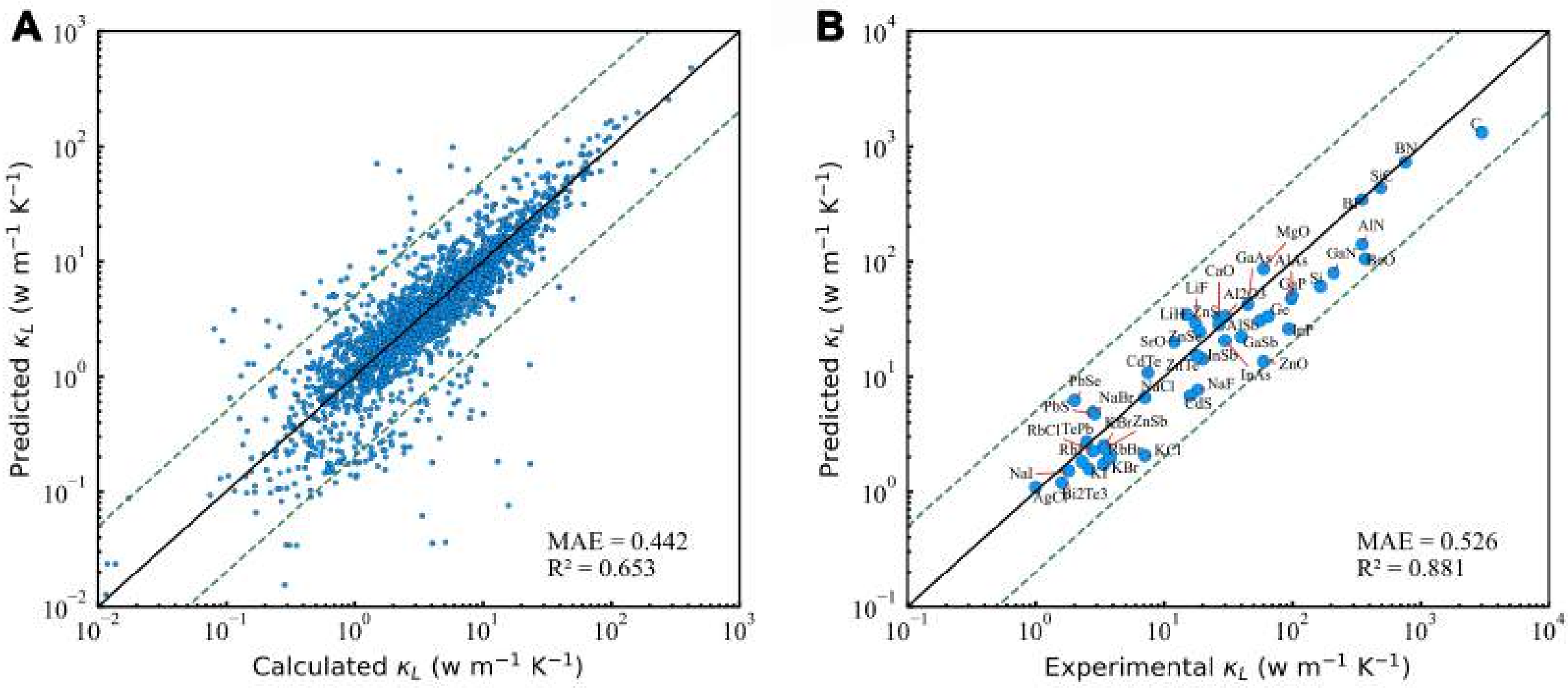

**Figure 6.** The comparison between $\kappa_L$ values predicted by PINK using (A) AFLOW[59] and (B) experimental[20,54-57] Grüneisen parameters $\kappa_L$ values at 300 K. Dashed lines indicate deviations within half an order of magnitude from reference values. PINK: Physical-informed kappa.

Even semi-empirical models, such as the Slack model, require time-consuming calculations or experimental data to determine the necessary input parameters, often taking several hours to complete. In contrast, PINK offers a significant advantage by predicting $\kappa_L$ and related physical properties directly from a CIF file in just a few seconds, regardless of the material's complexity.

While traditional methods such as PBTE and Green-Kubo can perform efficiently for single-element systems, their computational cost increases exponentially with the number of atoms in the primitive cell, especially when calculating force constants[61,62]. PINK, which leverages the CGCNN model in combination with the Slack approximation, provides a highly efficient solution. This allows for rapid pre-screening of complex binary, ternary, and quaternary systems. As a result, PINK is an invaluable tool for identifying materials with high or low $\kappa_L$.

### High-throughput screening

The detailed workflow for high-throughput screening using empirical calculations is illustrated in Figure 7. The screening process commenced with 377,221 compounds sourced from the Materials Discovery Database. Initial predictions of $\kappa_L$ were made for these compounds, along with bulk modulus (GPa), shear modulus (GPa), transverse and longitudinal wave sound velocities (m/s), speed of sound (m/s), Poisson's ratio ($\nu$), Grüneisen parameter ($\gamma$), acoustic Debye temperature ($\theta_a$, K), and $\kappa_L$ (W/m·K) can be downloaded at the link: https://github.com/Jack-Liu0227/AI4Kappa/tree/master/JMI_Supporting_Information.

To refine the dataset, preliminary screening criteria were applied. Since thermoelectric materials are semiconductors, band gaps were restricted to the range of 0.1-3.0 eV. To ensure stability, the energy above the convex hull was limited to zero or less[38]. This initial filtering reduced the dataset to 30,199 materials. Further exclusion of materials containing radioactive elements resulted in 26,305 candidates.

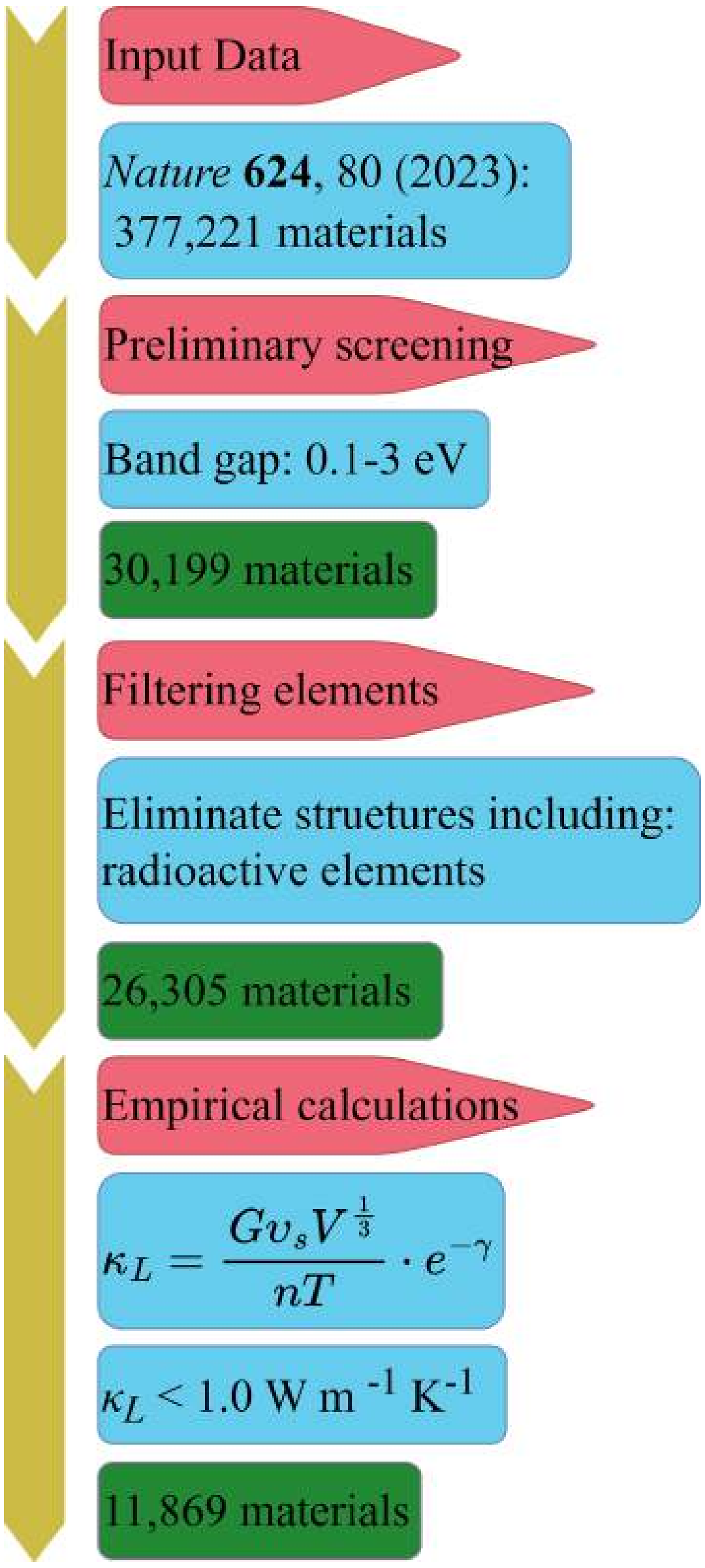

**Figure 7.** Flowchart of the high-throughput screening process, illustrating steps from data acquisition to filtering and empirical calculations for $\kappa_L$ prediction.

Subsequently, the CGCNN model was utilized to predict shear and bulk moduli, which were then employed to estimate $\kappa_L$ using the Slack model at 300 K. Materials with $\kappa_L$ values below 1 W·m$^{-1}$·K$^{-1}$ were identified as promising candidates for thermoelectric applications. This filtering yielded 11,869 materials, documented in Nature-filtered-low-Kappa.csv.

Additionally, using the Materials Project API, 54,359 structures with band gaps between 0.1-3.0 eV and no radioactive elements were extracted. PINK was employed to predict $\kappa_L$, resulting in a refined dataset of 21,001 low $\kappa_L$ materials, detailed in MP-semiconductor-low-kappa.csv.

### Statistical analysis of screening results

The process of screening 11,869 materials with low $\kappa_L$ values ($\kappa_L \leq 1$ W·m$^{-1}$·K$^{-1}$) has been detailed in a separate CSV file (Nature-filtered-low-kappa.csv), which also includes the associated material data. Statistical results for materials that have passed this screening are shown in Figure 8. The histogram in Figure 8A shows the distribution of $\kappa_L$ values, with the majority between 0.1 and 0.5, highlighting a promising subset of high-performance thermoelectric materials. Meanwhile, Figure 8B represents the distribution of crystal structures among these screened materials, specifying the number of space groups for cubic systems. The analysis reveals an inverse trend between symmetry and material count - fewer materials are found with higher symmetry, as in cubic systems (262 materials), while lower symmetry, such as in triclinic systems, is associated with a larger count. For cubic structures, the relevant space group numbers include 198, 205, 214, 215, 216, 217, 225, 227, 229, and 230. Low $\kappa_L$ values are widely acknowledged as critical for improving the efficiency of thermoelectric materials in converting waste heat to electrical energy.

To analyze the compositional distribution of promising thermoelectric materials, a histogram was generated to display the elemental distribution within the screened materials, as shown in Figure 8C. This diagram underscores the importance of the 20 most common elements in compounds with $\kappa_L$ values less than 1. Cesium bromine, rubidium, and adenosine oxide emerge as the elements most frequently encountered. Additionally, elements such as oxygen (O) and selenium (Se) are also prevalent in materials with low $\kappa_L$ values. The corresponding electronegativity values of these elements are provided at the top of each column in Figure 7C. A thorough examination of the electronegativity data reveals that elements with higher electronegativity are more likely to form stronger ionic bonds. Among the ten elements that occur most frequently, the majority exhibit electronegativity values greater than 2.5. Interestingly, elements often associated with low thermal conductivity, such as cesium and selenium, are part of this group. Moreover, fluorine, characterized by its high electronegativity, readily forms ionic compounds with alkali metals, including cesium, rubidium, potassium, and sodium.

### First-principle validation

On the basis of the results from our high-throughput screening and prior experience, we observed that compounds containing heavy elements and Group VIA elements generally exhibit lower thermal conductivity. Given the structural feasibility and computational efficiency, we selected the cubic structure for our study. Consequently, $Ag_3Te_4W$ and $Ag_3Te_4Ta$ were chosen as validation targets. To calculate $\kappa_L$ for a given material with a specific structure, a series of DFT calculations are performed within the volume of the primitive cell. To validate the materials screened by the PINK, including those with low $\kappa_L$, we have selected $Ag_3Te_4W$ and $Ag_3Te_4Ta$ as case studies for detailed analysis. Both crystals belong to space group 215. As illustrated in Figure 9A, X (W, Ta) atoms are tetrahedrally coordinated by four Te atoms, while Ag atoms occupy interstitial sites between neighboring tetrahedra. Phonon spectra calculations for these materials [Figure 9B and C] reveal no imaginary frequencies, confirming their dynamic stability and theoretical viability.

The $\kappa_L$ values of these compounds were calculated using the 3-phonon (3ph) method. Notably, $Ag_3Te_4X$ (X = W, Ta) exhibits ultralow $\kappa_L$, comparable to benchmark thermoelectric materials such as PbQ (Q = Te, Se) and SnSe[46,63,64]. Figure 10A compares the temperature-dependent $\kappa_L$ of $Ag_3Te_4X$ with state-of-the-art systems including SnSe, $Tl_9SbTe_6$[62], and PbQ. At 300 K, $Ag_3Te_4W$ and $Ag_3Te_4Ta$ demonstrate $\kappa_L$ values of 0.267 and 0.478 W·m$^{-1}$·K$^{-1}$ , respectively. By comparison, PbTe, PbSe, SnSe, and $Tl_9SbTe_6$ exhibit $\kappa_L$ values of 2.3, 2.64, 0.62, and 0.143 W·m$^{-1}$·K$^{-1}$ at the same temperature, respectively [Table 2]. The exceptionally low $\kappa_L$ of $Ag_3Te_4X$ positions these materials as promising candidates for thermoelectric applications.

**Table 2. The predicted $\kappa_L$ values from PINK were compared with those computed using DFT at 300 K**

| Materials | *n* | *V* (Å$^3$) | *G* (GPa) | B (GPa) | $\upsilon_s$ (m·s$^{-1}$) | $\gamma$ | $\kappa_{PINK}$ (W·m$^{-1}$·K$^{-1}$) | $\kappa_{DFT}$ (W·m$^{-1}$·K$^{-1}$) |
|---|---|---|---|---|---|---|---|---|
| PbTe[46] | 2 | 69.9891 | 24 | 38 | 1,927.318 | 2.180[63] | 3.5914 | 2.300 (Experiment) |
| PbSe[46] | 2 | 59.0544 | 27 | 24 | 2,035.793 | 2.660 | 2.493 | 2.640 (Experiment) |
| SnSe[64] | 8 | 226.257 | 15 | 24 | 1,783.133 | 2.300 | 0.681 | 0.680 (Experiment) |
| $Ag_3Te_4Ta$ | 8 | 250.951 | 13.776 | 45.133 | 3,074.950 | 2.231 | 0.628 | 0.478 (DFT) |
| $Ag_3Te_4W$ | 8 | 248.111 | 12.406 | 42.316 | 2,939.400 | 2.276 | 0.507 | 0.267 (DFT) |

In this comparison, *n* stands for the number of atoms in the primitive cell, *V* is the volume of the primitive cell, and *G* and *B* are the shear modulus and bulk modulus from Materials Project database for PbTe, PbSe and SnSe or our CGCNN model for $Ag_3Te_4Ta$ and $Ag_3Te_4W$. Additionally, $\upsilon_s$ refers to the average speed of sound, as defined in Equation (5), while $\gamma$ signifies the Grüneisen parameter, as given in Equation (6). PINK: Physical-informed kappa; DFT: density functional theory; CGCNN: crystal graph convolutional neural network.

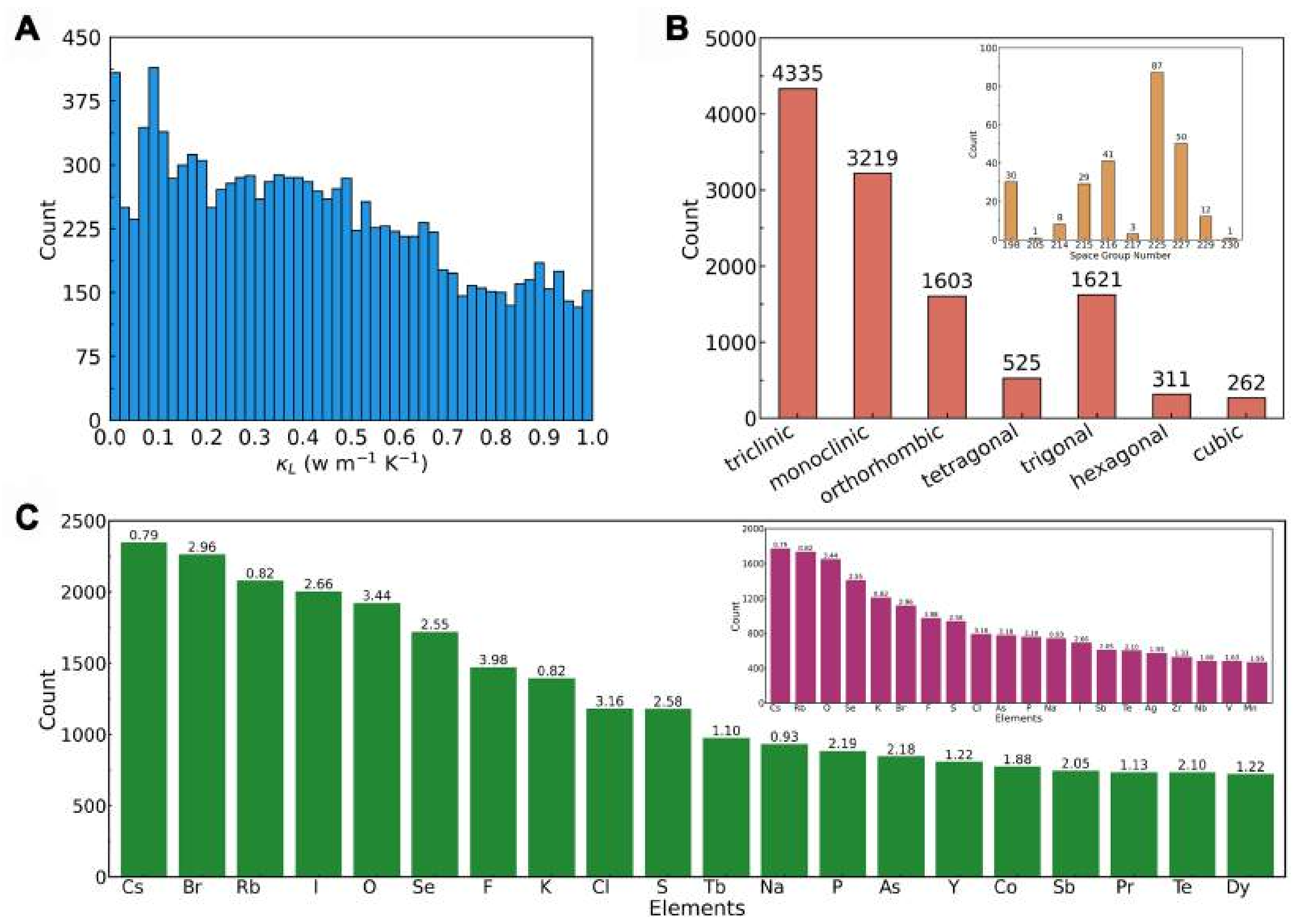

**Figure 8.** Statistical results of 11,869 screened candidates. (A) Distribution of $\kappa_L$ and corresponding count; (B) Distribution of crystal symmetry, with space group details for cubic symmetry shown in the inset; (C) Histogram of elemental distribution in 11,869 compounds, with electronegativity values indicated at the top of each column. The electronegativity results are as follows: Cs: 0.79, Br: 2.96, Rb: 0.82, I: 2.66, O: 3.44, Se: 2.55, F: 3.98, K: 0.82, Cl: 3.16, S: 2.58, Tb: 1.1, Na: 0.93, P: 2.19, As: 2.18, Y: 1.22, Co: 1.88, Sb: 2.05, Pr: 1.13, Te: 2.1, Dy: 1.22. The inset in the top-right corner is the counting number excluding lanthanide-containing materials.

Acoustic phonons typically serve as the dominant contributors to thermal transport in materials. As illustrated in Figures 9B and C and Figure 10B, acoustic phonon branches predominantly occupy low-frequency regimes, with low-frequency acoustic modes dominating the contribution to $\kappa_L$. To unravel the microscopic mechanisms underlying the ultralow $\kappa_L$, we systematically examined key parameters governing thermal conductivity - including heat capacity, phonon group velocities, weighted phase space, and scattering rates - for $Ag_3Te_4X$ (X = W, Ta) and $Tl_9SbTe_6$[62]. These analyses, presented in Figure 10B-F,

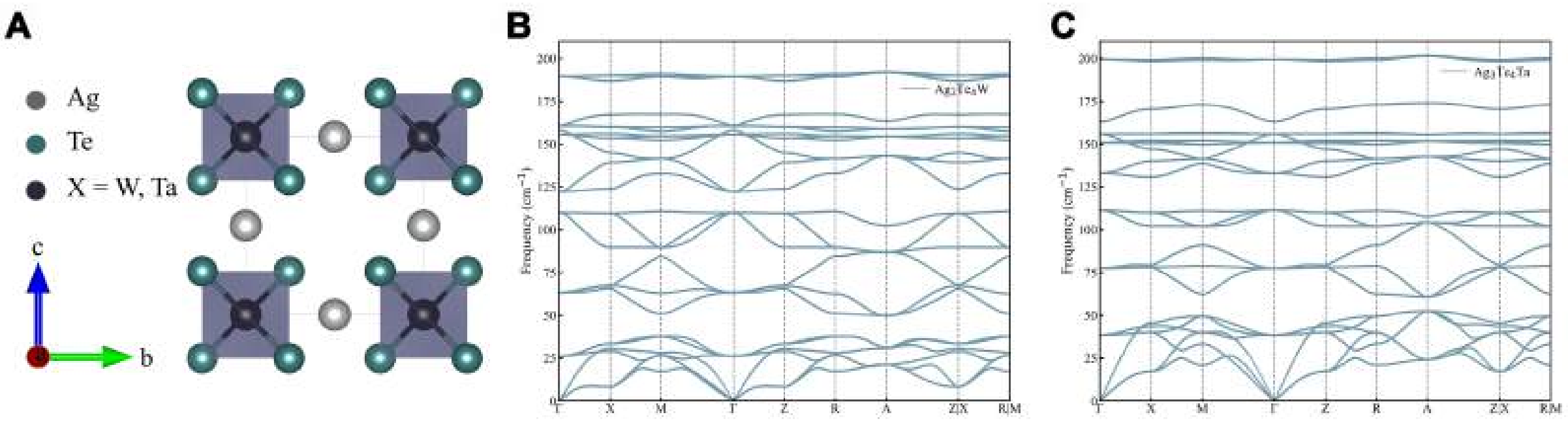

**Figure 9.** (A) The primitive crystal structures of $Ag_3Te_4X$ (X = W, Ta). Phonon dispersions for (B) $Ag_3Te_4W$ and (C) $Ag_4Se_4Ta$, respectively, along the high-symmetry points which are defined as Γ (0, 0, 0), X (0, 0.5, 0), M (0.5, 0.5, 0.0), Z (0, 0, 0.5), R (0, 0.5, 0.5), A (0.5, 0.5, 0.5).

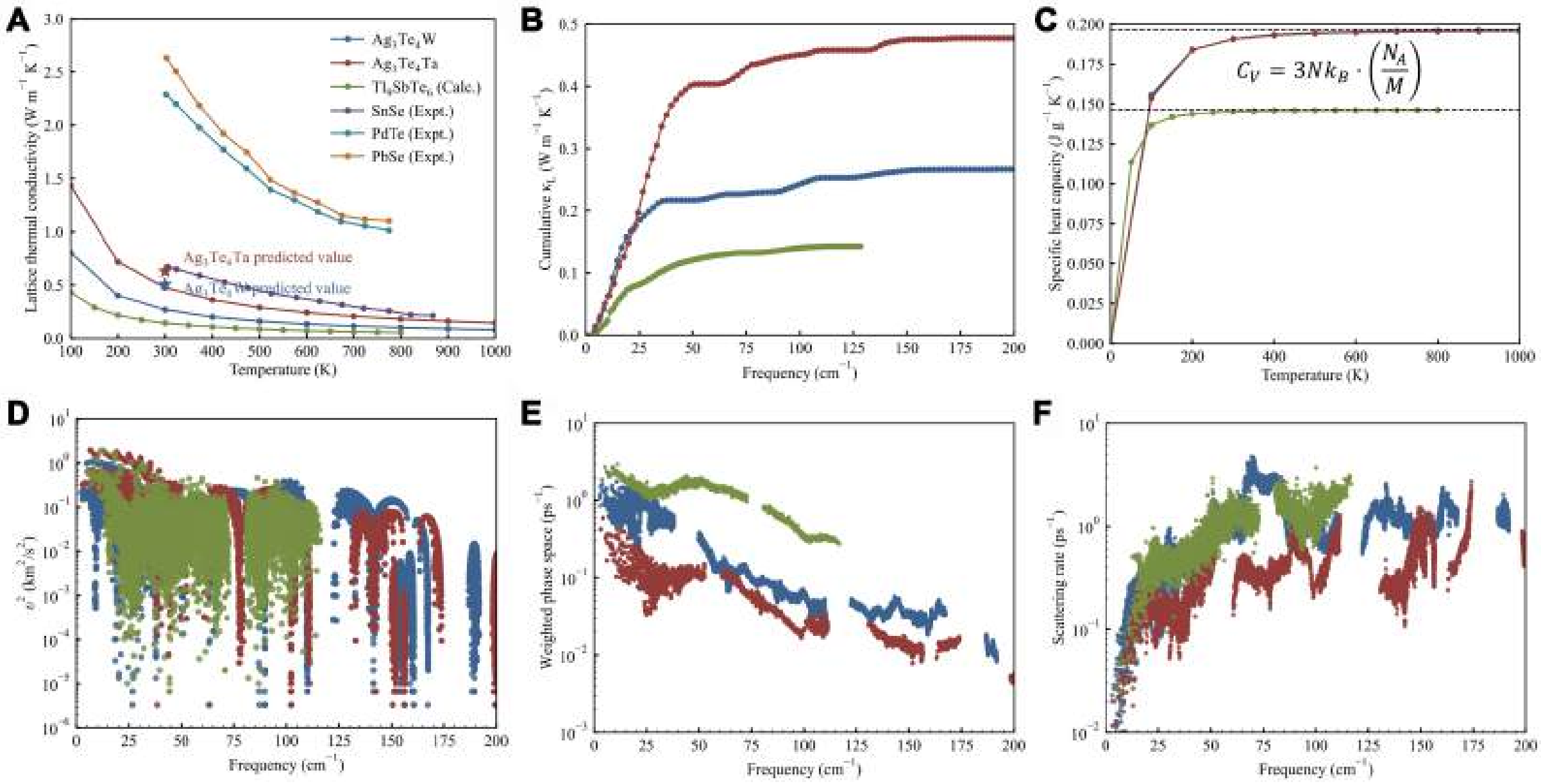

**Figure 10.** (A) $\kappa_L$ as a function of temperature for $Ag_3Te_4X$ (X = W, Ta), $Tl_9SbTe_6$[62], SnSe[46], and PbQ (Q = Te, Se)[46]. Comparison of microscopy heat transport parameters for $Ag_3Te_4X$ and $Tl_9SbTe_6$[62] at 300 K; (B) Cumulative $\kappa_L$ using 3ph methods; (C) Specific heat capacity ($C_V$) at constant volume; (D) Squared phonon group velocities ($\upsilon^2$) in the harmonic approximation; (E) Weighted phonon scattering phase space of 3ph; (F) Phonon scattering rates of 3ph.

provide critical insights into the interplay of phonon dynamics and thermal transport.

The specific heat capacity ($C_V$) of solids at elevated temperatures approximates the Dulong–Petit limit, defined as $3Nk_B$ ($N_A/M$), where $N$ denotes the number of atoms per formula unit, $k_B$ is the Boltzmann constant, $N_A$ the Avogadro constant, and $M$ the molar mass. DFT calculations yield $C_V$ values of 0.196, 0.197, and 0.146 $J \cdot g^{-1} \cdot K^{-1}$ for $Ag_3Te_4W$, $Ag_3Te_4Ta$, and $Tl_9SbTe_6$, respectively, closely aligning with Dulong–Petit predictions. Notably, the $C_V$ values for $Ag_3Te_4X$ (X = W, Ta) exceed those of $Tl_9SbTe_6$ and marginally surpass the 0.156 $J \cdot g^{-1} \cdot K^{-1}$ reported for PbTe[65].

Given the proportionality $\kappa_L \propto \upsilon^2$, we analyzed the frequency-dependent squared phonon group velocities ($\upsilon^2$) of $Ag_3Te_4X$ (X = W, Ta) and $Tl_9SbTe_6$, as illustrated in Figure 10D. Within the frequency range dominant for $\kappa_L$, $Ag_3Te_4Ta$ displays the highest $\upsilon^2$ (2 $km^2 \cdot s^{-2}$), an order of magnitude lower than PbTe's reported $\upsilon^2$ of 14 $km^2 \cdot s^{-2}$ ($\kappa_L$ = 2 $W \cdot m^{-1} \cdot K^{-1}$ at 300 K)[66]. $Ag_3Te_4W$ exhibits intermediate values, while $Tl_9SbTe_6$ shows the lowest $\upsilon^2$.

To elucidate phonon scattering mechanisms, we calculated the weighted phonon scattering phase space ($W_3$), which quantifies available phonon-phonon interaction pathways. As shown in Figure 10E, $Ag_3Te_4Ta$ has the smallest $W_3$, contrasting sharply with the largest $W_3$ observed in $Tl_9SbTe_6$. Similarly, phonon scattering rates [Figure 10F] are significantly higher for $Tl_9SbTe_6$ than for $Ag_3Te_4Ta$. These results collectively underpin the ultralow $\kappa_L$ of $Ag_3Te_4X$ (X = W, Ta).

### Discussions

In this work, we present PINK, a high-throughput computational framework designed to enhance the prediction of $\kappa_L$ across diverse materials. Building on this platform, several strategic directions emerge for future refinement and application. First, integrating PINK with experimental databases and materials informatics platforms could accelerate the discovery of novel materials for thermoelectrics, thermal management in microelectronics, and energy conversion systems. Coupling high-throughput predictions[67] with experimental validation would enable rapid identification of high-performance materials, narrowing the gap between computational insights and functional material synthesis. Additionally, synergizing PINK with tools such as BoltzTraP[68] and TransOpt[69] could enable concurrent optimization of thermal and electrical transport properties in semiconductors.

A second critical opportunity lies in the precise engineering of multifunctional materials. By investigating the interaction between $\kappa_L$, mechanical properties (e.g., strength, elasticity), and environmental stability, researchers could design materials that simultaneously achieve thermal, mechanical, and operational demands in sectors such as aerospace, renewable energy, and advanced electronics. Such integration of properties would advance applications requiring both efficient heat regulation and structural resilience.

## CONCLUSIONS

We have developed a high-throughput framework, packaged as an application named PINK, designed to rapidly predict the $\kappa_L$ of materials based on the CIF files. The material space for $\kappa_L$ was expanded significantly, increasing by an order of magnitude, through predictions for 377,221 newly reported materials[38]. Through high-throughput screening, several materials with ultralow $\kappa_L$ were identified, and their predictions were validated using first-principles calculations.

Although first-principle calculations of $\kappa_L$ require significant computational resources, especially for phonon spectra and third-order force constant matrices, there are existing databases related to phonons and $\kappa_L$. For instance, Togo developed an automated workflow interfaced with Phonopy, which calculated phonon spectra, density of states, entropy, and heat capacity for over 11,000 materials, creating a phonon database available at https://github.com/atztogo/phonondb/blob/main/mdr/phonondb/README.md. AFLOW[53], a comprehensive database, includes thermal property data for 5,664 materials, though this represents only a small fraction of the total material space. PINK addresses this gap by extending $\kappa_L$ predictions to hundreds of thousands of materials, with accuracy contingent on the performance of its embedded CGCNN for elastic modulus prediction.

To enhance prediction accuracy, future advancements could integrate advanced crystal graph convolutional networks. Examples include the orbital graph convolutional neural network (OGCNN), which considers orbital roles[70]; the materials graph network (MEGNet), incorporating outfield information[71]; the geometric-information-enhanced crystal graph neural network (GeoCGNN), which integrates topological and geometric structure data[72]; and the atomistic line graph neural network (ALIGNN), which includes bond angle details[73]. Other notable approaches include the graph-attention graph neural network (GATGNN), utilizing attention mechanisms[74], and the scalable global graph attention neural network model DeeperGATGNN, featuring differentiable group normalization (DGN) and skip connections[75]. Moreover, powerful descriptors such as SOAP[76] and Voronoi tessellations[77] could be employed to further elucidate the link between crystal structures and material properties. Ruff *et al.* introduced a connection-optimized crystal graph network (coGN/coNGN), leveraging message passing and line graph templates[78]. Their model demonstrated exceptional performance on the MatBench benchmark dataset[52], outperforming other models and establishing itself as the leading general-purpose model in the benchmark.

## DECLARATIONS

### Authors' contributions

Conceptualization, methodology, software, data curation, visualization, writing-original draft preparation: Liu, Y.

Writing-review and editing, supervision, project administration, funding acquisition: Gao, Z.

Performed data analysis and interpretation: Wang, X.; Hao, Y.; Li, X.

Investigation, discussion: Ding, X.; Lookman, T.; Sun, J.

### Availability of data and materials

All data are available at https://github.com/JackLiu0227/AI4Kappa/tree/master/JMI_Supporting_Information.

### Financial support and sponsorship

We acknowledge the support from the National Natural Science Foundation of China (No.12104356 and No.52250191). This work is sponsored by the Key Research and Development Program of the Ministry of Science and Technology (No.2023YFB4604100). We also acknowledge the support of the HPC Platform, Xi'an Jiaotong University.

### Conflicts of interest

Ding, X. is an Associate Editor of *Journal of Materials Informatics*. He was not involved in any steps of the editorial process, including reviewer selection, manuscript handling, or decision-making. The other authors declare that there are no conflicts of interest.

### Ethical approval and consent to participate

Not applicable.

### Consent for publication

Not applicable.